\documentclass[aps,prl,twocolumn,showpacs,superscriptaddress,nofootinbib,nolongbibliography]{revtex4-2}  % for review and submission
\usepackage{changes} 
\usepackage[para,online,flushleft]{threeparttable}
\usepackage{graphicx}  % needed for figures
\usepackage{dcolumn}   % needed for some tables
\usepackage{bm}        % for math
\usepackage{amssymb}   % for math
\usepackage{amssymb,amsmath,amsfonts} % for math
\usepackage[breaklinks=true,debug=true]{hyperref}
\usepackage{braket}    % for QM notation
\usepackage{multirow}
\usepackage{adjustbox}
\usepackage{gnuplottex}
\usepackage[T1]{fontenc}
\usepackage[utf8]{inputenc}
\usepackage{color}
\usepackage{dsfont}
\usepackage{tikz}
\usepackage{filecontents}
\usepackage{pgfplots}
\usepgfplotslibrary{fillbetween}
\usetikzlibrary{spy,shadows}
\usetikzlibrary{backgrounds}

\usepackage[para,online,flushleft]{threeparttable}

\usepgfplotslibrary{groupplots}
\pgfplotsset{compat=newest}
\usepgfplotslibrary{units}
\pgfdeclarelayer{background}
\pgfdeclarelayer{foreground}
\pgfsetlayers{background,main,foreground}

\begin{document}

%\widetext
%\leftline{Version 01 as of \today}
%\leftline{To be submitted to PRL}

\title{Constraining theoretical Corrections to Gamow-Teller Transition Rates}

\author{L.\,Xayavong} 
\email{xayavong.latsamy@yonsei.ac.kr}
\affiliation{Department of Physics, Yonsei University, Seoul 03722, South Korea}
\author{Y.\,Lim}
\email{ylim@yonsei.ac.kr}
\affiliation{Department of Physics, Yonsei University, Seoul 03722, South Korea}
%\author{N.\,A.\,Smirnova} 
%\email{smirnova@lp2ib.in2p3.fr}
%\affiliation{CENBG (CNRS/IN2P3 -- Université de Bordeaux), 33175 Gradignan cedex, France}
\vskip 0.25cm  
\date{\today}

\begin{abstract}  

 We propose two novel constraints for the theoretical corrections to Gamow-Teller transition rates. The first, derived from a two-level model, predicts forbidden regions within the plane defined by the isospin-mixing correction, $\delta_{C1}$, and the ratio, $\eta$, of the isospin-symmetry Gamow-Teller matrix elements between the upper and lower admixed states. 
 It serves as a filter for the theoretical calculations, particularly effective for small values of $|\eta|$. 
 The other employs experimental $ft$ values, incorporating the upper admixed states, and exploits mirror symmetry to eliminate isospin-invariant and nuclear structure-independent quantities. % presented in the master decay formula. 
 This approach not only offers an alternative mean for collectively testing theoretical corrections but also, as a byproduct, enables the extraction of $\eta^2$. This provides another sensitive test for the isospin-conserving component of nuclear Hamiltonian. 
 %, assuming that the overall contribution of all theoretical corrections does not exceed 50\,\%. 
 Our investigation reveals a substantial cancellation of among radiative correction contributions in these tests. 

\end{abstract}

%\pacs{21.60.Cs, 23.40.Bw, 23.40Hc, 27.30.+t}
\maketitle

%\section{Introduction}
{\it Introduction :} In the standard-model framework, the Gamow-Teller\,(GT) process is %exclusively 
governed by the axial-vector term of the weak current. In the non-relativistic limit, this term reduces to the operator $\bm{\sigma}\bm{\tau}_{\pm}$, allowing for a change of total angular momentum and isospin by one unit. Consequently, acting this operator on a specific initial state can result in several possible final states. Due to this characteristic, the theoretical description of GT transitions strongly depends on the chosen model of nuclear structure\,\cite{GT,BROWN1985347,Kumar2016,PhysRevC.105.034349,WILKINSON1974365,PhysRevC.7.930,WILKINSON1973470}. 
Notably, unlike the Fermi matrix elements which are
solely determined by the isospin quantum numbers, 
the GT matrix elements, even between states with definite isospin, are not analytically known. 
Understanding the impact of isospin-symmetry breaking on this process poses an even greater challenge, 
since isospin-nonconserving interactions constitute only a small portion of the complete nuclear Hamiltonian\,\cite{OrBr1985,Smirnova2003,Lam2013,XaNa2018,XaNa2022,xayavong2022higherorder,xayavong2023shellmodel}. \\
\indent 
The master formula for GT transitions is expressed as
\begin{equation}\label{ft}
\displaystyle ft(1+\delta_{R}')(1-\delta_{C} + \delta_{NS}) = \frac{K}{\mathcal{M}^2G_F^2q_A^2g_A^2V_{ud}^2(1+\Delta_R)},  
\end{equation}
where the left-hand side contains the experimental $ft$ value, which is the product of the statistical rate function\,($f$) and partial half-life\,($t$), 
along with theoretical corrections $\delta_C$, $\delta_R'$, and $\delta_{NS}$. These corrections account for isospin-symmetry breaking, transition-dependent, and nuclear structure-dependent radiative effects, respectively. On the right-hand side, apart from the isospin-symmetry GT matrix element, $\mathcal{M}$ and its quenching factor $q_A$ that compensates for the systematic deviation of shell-model predictions from experimentally deduced values\,\cite{BROWN1985347,GT}, only nucleus-independent quantities are present. These include $K$, a combination of fundamental constants\,\cite{HaTo2020}; $G_F$ and $g_A$, the Fermi\,\cite{MuLan,10.21468/SciPostPhysProc.5.016} and axial-vector coupling\,\cite{PhysRevLett.122.242501,PhysRevLett.129.232502} constants, respectively; $\Delta_R$, another radiative correction term that merely depends on a specific type of weak current% governing the transition under consideration
\,\cite{PhysRevD.103.113001,PhysRevD.108.053003,PhysRevD.100.013001}; and $V_{ud}$, the top-left element of the Cabibbo–Kobayashi–Maskawa quark-mixing matrix\,\cite{HaTo2020,PhysRevLett.10.531,10.1143/PTP.49.652}. When considering a pair of mirror GT transitions, all universal constants and isospin-invariant quantities in Eq.\,\eqref{ft} can be eliminated. \\
The obtained equation at first order is, 
\begin{equation}\label{test1}
 \Delta_I \approx \frac{ft^+}{ft^-} - (\delta_C^+ - \delta_C^-) + (\delta_{NS}^+ - \delta_{NS}^-) + (\delta_R'^+ - \delta_R'^-) - 1 , 
\end{equation}
where the label `$+$'(`$-$') correspond to the $\beta^+$($\beta^-$) emission. Given the expectation of a zero residual ($\Delta_I=0$), 
Eq.\,\eqref{test1} has been used as a collective test for theoretical corrections 
on its right-hand side, employing the experimental data on $ft^+/ft^-$. A notable agreement was achieved between theoretical predictions and experimental data in the previous shell-model studies\,\cite{Smirnova2003,xayavong2023shellmodel}, although $(\delta_{NS}^+-\delta_{NS}^-)=0$ and $(\delta_R'^+-\delta_R'^-)=0$ were assumed. 
In general, both of these radiative corrections consist of a small isovector component\,\cite{ToHa2002,TOWNER1992478} that could contribute to a mirror asymmetry. 
%In general, both of these radiative corrections consist of a small isovector contribution, resulting in a variation under the mirror symmetry\,\cite{ToHa2002}. 
%, which might not be justified at the current experimental precision level. 
%\added{YH : is it not justfied or justfied? If the current level of precision is not able to distinguish them, it can be justified. } 
%\tcr{LX: According to our current result given in the top-left panel of Fig.\,\ref{fig2}, $\Delta_I$ is not perfectly consistent with zero since $\chi^2/\nu\gg1$, even though its average is very close to zero. This discrepancy might suggest potential underestimate of uncertainties in the isospin-symmetry breaking correction and/or experimental $ft$ values, or highlights the significance of the missing radiative corrections. } 
A finite value of $\Delta_I$ could also be attributed to the existence of second-class weak currents\,\cite{PhysRev.112.1375,PhysRev.132.738,DELORME1971317,Wilkinson2000}, which are absent in the standard model. $\Delta_I$ beyond standard model, however, seems to be negligible compared to current uncertainties in nuclear structure calculations\,\cite{PhysRevLett.130.192502,PhysRevLett.128.202502,HaTo2020,xayavong2023shellmodel}. 
\indent
Within the shell-model framework, $\delta_C$ can be decomposed into two components:  
\begin{equation}
    \delta_C\approx\delta_{C1} + \delta_{C2}, 
\end{equation}
where $\delta_{C1}$ accounts for configuration admixtures within the shell-model valence space induced by the isospin-nonconserving part of the effective Hamiltonian. Whereas, $\delta_{C2}$ accounts for the mismatch between proton and neutron radial wave functions, compensating for the isospin admixtures extending beyond the shell-model valence space. The interference terms are of higher orders. For additional details, please refer to Refs.\,\cite{xayavong2023shellmodel,xayavong2022higherorder}. 
\\
\indent
The purpose of this Letter is to introduce two novel constraints for theoretical corrections, offering a greater selectivity compared to Eq.\,\eqref{test1}. 
One of them is exclusively sensitive to $\delta_{C1}$ and relies only on predictions of a two-level model of isospin mixing\,\cite{xayavong2023shellmodel}, without the need for experimental data. This method can be expected to work at the weak isospin-mixing limit. 
The other, although originated from the master formula\,\eqref{ft}, differs from Eq.\,\eqref{test1} by incorporating GT transitions to second admixed states. This introduces a novel collective test for theoretical corrections. 
Additionally, the contributions of the radiative corrections are empirically investigated based on the existing results obtained for superallowed $0^+\rightarrow0^+$ Fermi $\beta$ decay\,\cite{ToHa2002,TOWNER1992478}. It is also demonstrated that the inclusion of an upper pair of mirror GT transitions allows for an experimental determination of $\eta^2$, serving as a test, which is solely dependent on the isospin-symmetry part of the nuclear Hamiltonian. 
%\section{Two-level mixing constraint for $\delta_{C1}$}
\\
\begin{center}
\begin{figure*}[ht!]
\includegraphics[]{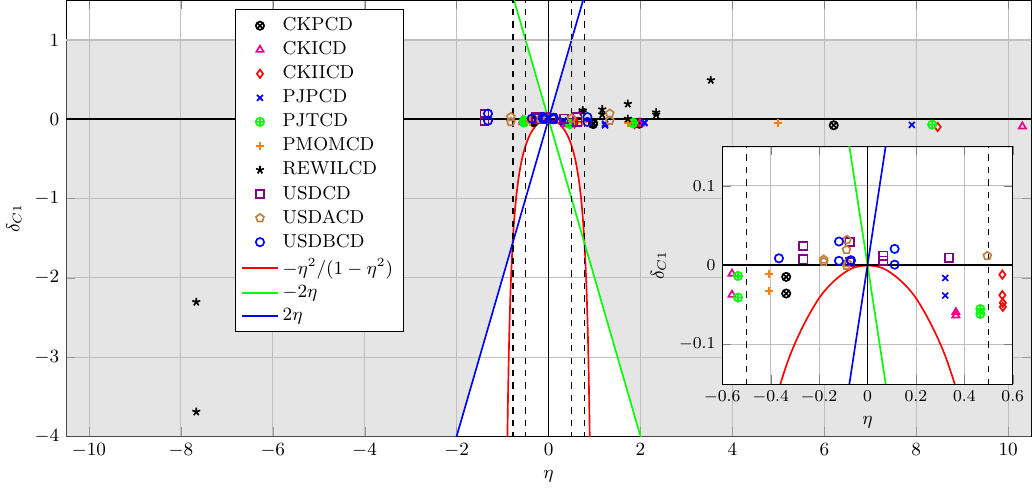}
\caption{The shaded areas illustrate the physical admissible regions of $\delta_{C1}$ based on the predictions from the two-level model. The points represent shell-model values obtained with various effective interactions, mainly taken from Ref.\,\cite{xayavong2023shellmodel}. Additionally, data for GT transitions to excited states, not covered in Ref.\,\cite{xayavong2023shellmodel}, are calculated with the same model spaces and effective Hamiltonian. 
%%
%\added{ 
%The corresponding numerical data are provided in the supplementary material. 
%}
%%
} 
\label{fig1}
\end{figure*}
\end{center}
\indent
{\it Formalism:} While an exact shell-model calculation of $\delta_{C1}$ for a GT transition is extremely complicated, 
certain fundamental properties of this correction term can be identified using a fully analytic two-level mixing model\,\cite{xayavong2023shellmodel}. 
%\added{YH : I'm still wondering if it is not necessary to mention that $\mathcal{M}$ is assumed to be real.*-} 
%\tcr{LX: } 
Specifically, it describes $\delta_{C1}$ as a function of the isospin-mixing amplitude, $\alpha$, namely 
\begin{equation}\label{C1}
\displaystyle \delta_{C1} = -2\eta\alpha + \left( 1-\eta^2 \right) \alpha^2 + \mathcal{O}(\alpha^3), 
\end{equation}
where $\eta=\mathcal{M}_1/\mathcal{M}_0$, representing the ratio of the isospin-symmetry GT matrix elements between the upper\,(labeled as 1) and the lower (labeled as 0) admixed states. Note that the standard definition of $\delta_{C1}$, namely $\delta_{C1} = 1-(M/\mathcal{M}_0)^2$ where $M$ denotes the exact GT matrix element for the lower admixed state\footnote{In the two-level mixing model, this matrix element can be expressed as $M = \sqrt{1-\alpha^2}\mathcal{M}_0 + \alpha\mathcal{M}_1$.}, is utilized in deriving Eq.\,\eqref{C1}. For simplicity, we also assume that $|\alpha|$ is sufficiently small to make the terms beyond the second order in Eq.\,\eqref{C1} negligible. According to first-order perturbation theory, $\alpha$ is inversely proportional to the energy separation between admixed states. Consequently, couplings of the lowest state with those above the second state diminish progressively, justifying the use of the two-level mixing model. While it is also possible to derive a similar $\delta_{C1}$ formula for the upper admixed state\,(the second state), the justification for the two-level mixing might not hold due to its significant coupling to the next excited state\footnote{In general, the density of states increases with excitation energy. Therefore, the energy separation between the first and second states is generally larger than that between the second and third states.}. Hence, our study based on this simplified model exclusively concentrates on $\delta_{C1}$ for GT transitions to the lowest state. It is important to emphasize that Eq.\,\eqref{C1} predicts essential characteristics of $\delta_{C1}$ specific to GT transitions. One of these is that $\delta_{C1}$ can be positive or negative, depending on both sign and magnitude of %the parameters
$\alpha$ and $\eta$. A negative $\delta_{C1}$ results in increases in %the absolute value of
$|M|$ due to the dominant constructive contribution from the upper admixed state. Conversely, a positive $\delta_{C1}$ produces the opposite effect. Another interesting observation arises when $\eta=0$, as in cases where the isospin quantum number of the upper admixed state differs from that of the initial state by more than one unit. 
In this situation, we obtain $\delta_{C1}=\alpha^2$, akin to what is observed in Fermi transitions\,\cite{ToHa2008}. In addition, the presence of the first-order term in $\alpha$ in Eq.\,\eqref{C1} implies that $\alpha$ must be a real number. %, given that $\eta$ being a real number. 
Roots of the quadratic equation\,\eqref{C1} take the forms, 
\begin{equation}\label{alpha}
\alpha = \kappa\left( 1 \pm \sqrt{ 1 + \frac{\delta_{C1}}{\eta \kappa} } \right),
\end{equation}
where $\kappa=\eta/(1-\eta^2)$. The requirement of $\Im(\alpha)=0$ implies that the expression inside the square root in Eq.\,\eqref{alpha} must be positive, leading to the following boundaries,  
\begin{equation}\label{con1}
%\left\{
%\begin{array}{l}
\displaystyle -\frac{\eta^2}{1-\eta^2} \le \delta_{C1} \le 1, \hspace{0.1in} \text{if} \hspace{0.1in} |\eta|<1.   %\\[0.14in]
%\delta_{C1} \le \displaystyle \frac{\eta^2}{|1-\eta^2|},   \hspace{0.1in} \text{otherwise} .
%\end{array}
%\right.
\end{equation}
%and any $\delta_{C1}$ values are allowed if $|\eta|\ge1$. 
%%
The sign in front of the square root in Eq.\,\eqref{alpha} can be fixed by applying $0\le\alpha^2\le1$, as $\alpha$ represents a probability amplitude. Taking the square to Eq.\,\eqref{alpha} yields, 
\begin{equation}\label{a2}
\displaystyle \alpha^2 = \kappa^2 \left[ 2+\frac{\delta_{C1}}{\eta\kappa} \pm \left( 2 + \frac{\delta_{C1}}{\eta\kappa} - \frac{\delta_{C1}^2}{4\eta^2\kappa^2} \right)  \right] + \mathcal{O}(\delta_{C1}^3). 
\end{equation}
It can be shown that the positive sign imposes restrictions on certain values of $\eta$ without theoretical justification. Therefore, it is disregarded. %On the other hand, t
The negative sign results in the following boundaries, definite for all $\eta$ values, 
\begin{equation}\label{con2} 
  -2\eta\le\delta_{C1}\le2\eta . 
\end{equation}
Note also that $\delta_{C1}\le1$ holds by definition, irrespective of $\eta$. The combination of Eqs.\,\eqref{con1} and \eqref{con2} results in, 
\begin{equation}\label{con}
\left\{
\begin{array}{ll}
\displaystyle -\frac{\eta^2}{1-\eta^2} \le\delta_{C1}\le 2\eta, & \text{if}  \hspace{0.1in} \displaystyle |\eta|< \frac{1}{2} \\[0.14in]
\displaystyle -\frac{\eta^2}{1-\eta^2} \le\delta_{C1}\le 1, & \text{if}  \hspace{0.1in} \displaystyle \frac{1}{2} \le|\eta|< \frac{1}{4}\left(\sqrt{17}-1\right)  \\[0.14in]
-2\eta\le\delta_{C1}\le 1, & \text{otherwise}. 
\end{array}
\right.
\end{equation} 
It should be noticed that Eq.\eqref{con} yields $\delta_{C1}$ equals 0 at $\eta=0$. This conflicts with the expectation from Eq.\eqref{C1}, the starting point of our analysis, where we determine $\delta_{C1}=\alpha^2$ for $\eta=0$, as in Fermi transitions% as previously mentioned
. Hence, it is logical to state that the error in Eq.\eqref{con} is, at least, of order $\mathcal{O}(\alpha^2)$. A finer precision might be available if the higher-order terms are included %after 
in the Taylor expansion% of the $\alpha^2$ formula
\,\eqref{a2}. The physically admissible regions in the $(\eta,\delta_{C1})$ plane defined by Eq.\,\eqref{con}, are represented as the shaded areas in Fig.\,\ref{fig1}. 
\\
\begin{center}
\begin{figure*}[ht!]
\includegraphics[]{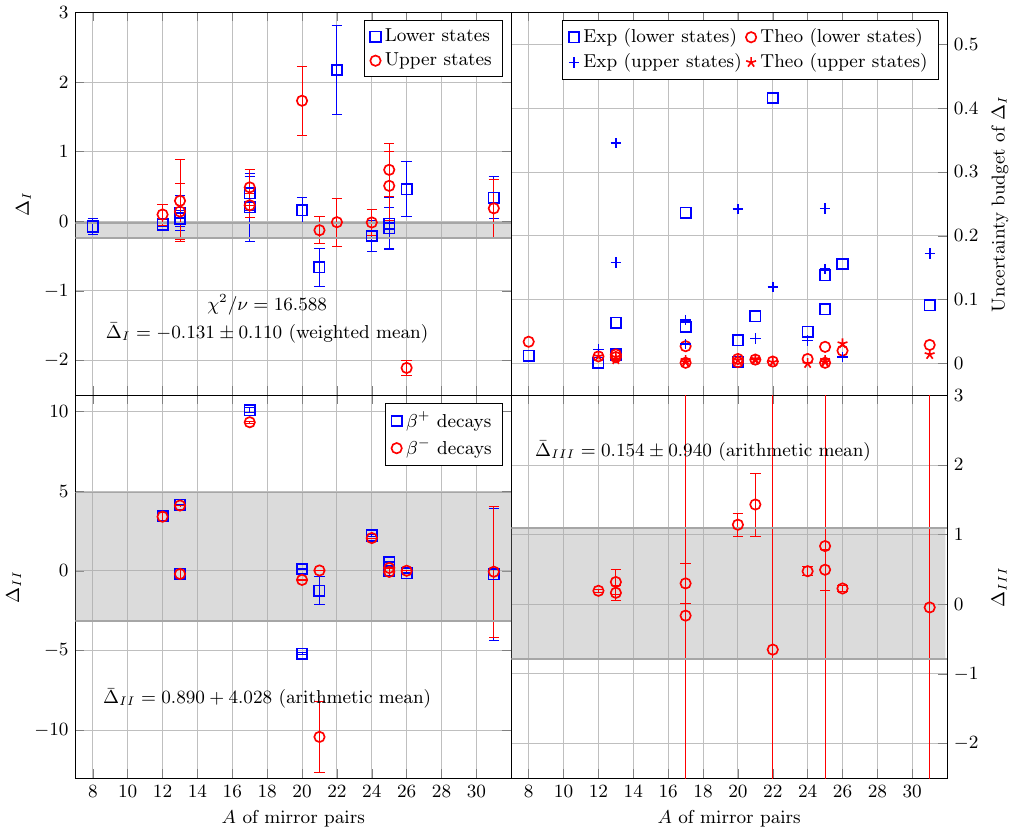}
\caption{\label{fig2}(Color online) Illustration of the shell-model tests for the residuals defined in Eq.\,\eqref{test1}, \eqref{test2}, and \eqref{test3}. The gray bands indicate the averages. The top-right panel displays the individual contributions to uncertainties of $\Delta_I$. 
The experimental inputs ($ft$ values) are taken from Ref.\,\cite{NNDCENSDF}. 
%The corresponding numerical data are provided in the supplementary material. 
}
\end{figure*}
\end{center}
\indent
To demonstrate the applicability of the boundaries\,\eqref{con} as a means for testing theoretical calculations of $\delta_{C1}$, we incorporate the shell-model results for the sixteen pairs of mirror GT transitions in the $p$ and $sd$ shells given in Ref.\,\cite{xayavong2023shellmodel} into Fig.\,\ref{fig1}. Additional data for %a few 
GT transitions to excited states, not considered in Ref.\,\cite{xayavong2023shellmodel} have been calculated within a similar approach. These theoretical data were obtained using exact diagonalizations of well-established effective isospin-nonconserving Hamiltonians and their corresponding isospin-symmetry versions for a selected model space. %The complete numerical data are available in the supplementary material. 
It is seen from Fig.\,\ref{fig1} that all the shell-model values are located inside the admissible regions, even under extreme scenarios where the two-level mixing model is not expected to be valid. Notably, the two large negative values near the bottom-left corner of the graph correspond to the decays of $^{22}$O($\beta^-$)$^{22}$F and  $^{22}$Si($\beta^+$)$^{22}$Al, where an exceptionally strong isospin mixing has been experimentally observed\,\cite{PhysRevLett.125.192503}. This test is particularly precise when $\eta$ has a small magnitude, narrowing admissible regions. 
\\ 
%%
%\section{New experimental constraint I} 
\indent
To establish a new experimental constraint, we compare the master formula, Eq.\,\eqref{ft}, between GT transitions to the first and second admixed states. In this way, all nuclear structure-independent quantities on the right-hand side of Eq.\,\eqref{ft} can be eliminated. The resulting expression reads 
\begin{equation}\label{test2}
\begin{array}{ll}
   \Delta_{II} & \displaystyle \approx \eta^2 - \frac{ft^0}{ft^1} \left[ 1+(\delta_{R}'^0-\delta_{R}'^1)+(\delta_{NS}^0-\delta_{NS}^1) \right. \\[0.15in]
   & \displaystyle-\left. (\delta_C^0-\delta_C^1) \right].
\end{array}
\end{equation} 
It is observed that Eq.\,\eqref{test2} is analogous to Eq.\,\eqref{test1}; they differ only 
by the interchange of the labels $0(1)$ with $+(-)$ and 
the presence of $\eta$ in the former, which arises from the lack of analogy between lower and upper admixed states. In principle, `0' and `1' could be any pair of excited states, given the availability of their corresponding $ft$ values. Since $\Delta_{II}=0$ is expected% to be zero
, Eq.\,\eqref{test2} serves as an alternative collective test for the theoretical corrections, including $\delta_{R}'^0$, $\delta_{R}'^1$, $\delta_{NS}^0$, $\delta_{NS}^1$, $\delta_C^0$ and $\delta_C^1$, together with $\eta^2$, using experimental data on $ft^0/ft^1$. 
The requirement of $\Im(\eta)=0$ also implies an additional constraint among these theoretical corrections:  
\begin{equation}\label{rr}
 (\delta_C^0-\delta_C^1) \le 1 + (\delta_{R}'^0-\delta_{R}'^1)+(\delta_{NS}^0-\delta_{NS}^1).   
\end{equation} 
With $(\delta_{R}'^0-\delta_{R}'^1)+(\delta_{NS}^0-\delta_{NS}^1)=0$, our shell-model results meet the condition\,\eqref{rr} except for 
$^{21}$F($\beta^-$)$^{21}$Ne and $^{21}$Mg($\beta^+$)$^{21}$Na, decaying to the second $\frac{5}{2}^+$ states. 
Our calculation for this exceptional case yields $\mathcal{M}_1\approx0$ which leads to a divergence in $\delta_C$\footnote{The leading terms of the correction are inversely proportional to the isospin-symmetry GT matrix element. See, for example, Ref.\,\cite{xayavong2023shellmodel} }. These transitions are thus excluded in the following studies. 
Among the remaining transitions, the largest $(\delta_C^0-\delta_C^1)$ values are approximately 55\,\%
corresponding to the decays of $^{17}$Ne($\beta^+$)$^{17}$F and $^{26}$Na($\beta^-$)$^{26}$Mg. 
Furthermore, Eq.\,\eqref{test2} enables the extraction of $\eta^2$ when the contribution from the corrective terms on its right-hand side are insignificant. Essentially, $\eta^2$ solely depends on the isospin-symmetry component of the nuclear Hamiltonian. 
%Additionally, Eq.\,\eqref{test2} could potentially enable an extraction of $\eta^2$, which is solely dependent on the isospin-symmetry component of the nuclear Hamiltonian, particularly when the contribution from the corrective terms on its right-hand side are insignificant. 
The comparison between these experimentally extracted and shell-model values of $\eta^2$ is given in Table\,\ref{tab1}. While both datasets are found to be consistent with the boundaries\,\eqref{con} and the shell-model values for $\delta_{C1}$, a substantial discrepancy emerges between them in the majority of cases. 
\\
\indent
According to previous studies on the superallowed $0^+\rightarrow0^+$ Fermi $\beta$ decay\,\cite{ToHa2002,ToHa2008}, $\delta_R'$ comprises two components: one dependent on the atomic number, $Z$ and the other on the transition $Q$ value. 
The $Z$-dependent component is common for both transitions to lower and upper admixed states, thus not contributing to $\Delta_{II}$. Additionally, the universal Born-graph contribution to $\delta_{NS}$ should cancel out identically in Eq.\,\eqref{test2}. 
This latter cancellation also applies to $\Delta_I$ in Eq.\,\eqref{test1}. 
%%
%\section{New experimental constraint II} 
%%  
 \begin{table*}[ht!]
 \caption{Comparison of $\eta^2$ between the values extracted from experimental $ft^0/ft^1$ data via Eq.\,\eqref{test2} neglecting the theoretical corrections, and those obtained from shell-model calculations. The experimental $\eta^2$ values are averaged between $\beta^+$ transitions and their $\beta^-$ mirror partners. 
 Columns five and six list the shell-model values for $\delta_C^0-\delta_C^1$, expressed as percentages, which are to be validated with Eq.\eqref{rr}. 
 } 
 \label{tab1} 
 \begin{threeparttable}
 \begin{ruledtabular}
 \begin{tabular}{c|c|c|c|c|c|c} 
Mirror pairs &  $J_i^\pi T_i$ & $J_f^\pi T_f$\tnote{c}	&	$\eta^2_{exp}$	&	$\eta^2_{theo}$	& $\delta_C^0-\delta_C^1$ ($\beta^+$) & $\delta_C^0-\delta_C^1$ ($\beta^-$) \\
\hline
$^8$Li$(\beta^-)^8$Be; $^8$B($\beta^+$)$^8$Be & $2^+1$ & $2^+0$	&	409.261$\pm$20.723	&	56.454$\pm$27.235\tnote{a} & 0.674$\pm$2.622	& -12.556$\pm$3.194 \\
$^{12}$B($\beta^-$)$^{12}$C; $^{12}$N($\beta^+$)$^{12}$C	& $1^+1$ & $0^+0$ &	0.308$\pm$0.004	&	3.722$\pm$0.285\tnote{a} & 8.413$\pm$1.234	& -5.545$\pm$0.901\\
\hline
$^{13}$B($\beta^-$)$^{13}$C; $^{13}$O($\beta^+$)$^{13}$N	& $\frac{3}{2}^-\frac{3}{2}$ & $\frac{1}{2}^-\frac{1}{2}$ &	0.271$\pm$0.026		&	0.110$\pm$0.320	& -5.396$\pm$0.419 & -9.409$\pm$1.044 \\
& $\frac{3}{2}^-\frac{3}{2}$ & $\frac{3}{2}^-\frac{1}{2}$ &	0.302$\pm$0.054	&	4.419$\pm$0.400\tnote{a} & 2.820$\pm$1.522 & -2.158$\pm$0.659 \\
\hline
$^{17}$N($\beta^-$)$^{17}$O; $^{17}$Ne($\beta^+$)$^{17}$F	& $\frac{1}{2}^-\frac{3}{2}$ & $\frac{3}{2}^-\frac{1}{2}$ &	4.626$\pm$0.139	&	12.460\tnote{a}		& 57.994$\pm$2.633 & 14.344$\pm$0.694 \\
& $\frac{1}{2}^-\frac{3}{2}$ & $\frac{1}{2}^-\frac{1}{2}$ &	444.048$\pm$62.749	&	12.429\tnote{a}	& -2.259$\pm$0.160 & 8.124$\pm$0.117 \\
\hline
$^{20}$F($\beta^-$)$^{20}$Ne; $^{20}$Na($\beta^+$)$^{20}$Ne	& $2^+1$  & $2^+0$ &	6.180$\pm$0.047	&	1.361\tnote{a}	& -5.963$\pm$0.682 & -14.205$\pm$0.248 \\
$^{20}$O($\beta^-$)$^{20}$F; $^{20}$Mg($\beta^+$)$^{20}$Na	& $0^+2$ & $1^+1$ &	0.869$\pm$0.058	&	0.551	& 23.498$\pm$0.532 & 10.011$\pm$0.138 \\
\hline 
$^{21}$F($\beta^-$)$^{21}$Ne; $^{21}$Mg($\beta^+$)$^{21}$Na	& $\frac{5}{2}^+ \frac{3}{2}$ & $\frac{3}{2}^+\frac{1}{2}$ &	9.651$\pm$1.345	&	5.479	& -11.661$\pm$0.736 & -12.463$\pm$0.239 \\
& $\frac{5}{2}^+ \frac{3}{2}$ & $\frac{5}{2}^+\frac{1}{2}$ &	16.764$\pm$5.383	&	0.017\tnote{a}	& 428.109$\pm$0.654\tnote{b} & 262.784$\pm$0.350\tnote{b} \\
\hline 
$^{22}$O($\beta^-$)$^{22}$F; $^{22}$Si($\beta^+$)$^{22}$Al	& $0^+ 3$ & $1^+2$ &	11.484$\pm$1.346	&	58.757\tnote{a}	& 0.922$\pm$0.144 & 17.693$\pm$0.324 \\
$^{24}$Ne($\beta^-$)$^{24}$Na; $^{24}$Si($\beta^+$)$^{24}$Al	& $0^+ 2$ & $1^+1$ &	0.866$\pm$0.030	&	2.976\tnote{a}	& 6.471$\pm$0.498 & 0.098$\pm$0.428 \\
\hline

$^{25}$Na($\beta^-$)$^{25}$Mg; $^{25}$Si($\beta^+$)$^{25}$Al	& $\frac{5}{2}^+ \frac{3}{2}$  & $\frac{5}{2}^+\frac{1}{2}$ &	0.133$\pm$0.019	&	0.111$\pm$0.307	& -3.560$\pm$2.541 & -7.652$\pm$0.835 \\
& $\frac{5}{2}^+\frac{3}{2}$ & $\frac{3}{2}^+\frac{1}{2}$ &	0.560$\pm$0.048	&	0.913$\pm$0.355	& 9.096$\pm$0.223 & -0.919$\pm$0.017 \\
\hline
$^{26}$Na($\beta^-$)$^{26}$Mg; $^{26}$P($\beta^+$)$^{26}$Si	& $3^+ 2$ & $2^+1$ &	0.031$\pm$0.010	&	0.006$\pm$0.004	& -48.954$\pm$3.456 & 54.833$\pm$1.196 \\
$^{31}$Al($\beta^-$)$^{31}$Si; $^{31}$Ar($\beta^+$)$^{31}$Cl	& $\frac{5}{2}^+ \frac{5}{2}$ & $\frac{3}{2}^+\frac{3}{2}$ &	2.192$\pm$0.222	&	1.785$\pm$4.123	& 16.739$\pm$2.965 & 7.789$\pm$1.273 \\
 \end{tabular}
 \end{ruledtabular}
 \begin{tablenotes}
{ \raggedright
%\item[a] Additional information, such as additional quantum numbers specifying the initial and final states, is available in the supplementary material. \\
\item[a] Disagree within conservative 50\,\% experimental error. \\
\item[b] According to Eq.\,\eqref{rr}, these values, exceeding 100\%, would be unphysical if $(\delta_{R}'^0-\delta_{R}'^1)+(\delta_{NS}^0-\delta_{NS}^1)\approx0$. \\
\item[c] The isospin quantum numbers for the first and second admixed states are identical for all cases. \\
}
\end{tablenotes}
\end{threeparttable}
\end{table*}
\indent
It is clearly seen from Table\,\ref{tab1} that, despite allowing for a conservative 50\,\% uncertainty to cover the remaining contribution of the missing radiative corrections, significant discrepancies between the experimental $\eta^2$ values and the shell-model predictions persist. Therefore, it is likely that $\eta^2$ is the largest uncertainty source in Eq.\,\eqref{test2}. In order to eliminate this isospin-invariant parameter from Eq.\,\eqref{test2}, we combine two pairs of mirror GT transitions, leading to   
\begin{equation}\label{test3}
\begin{array}{ll}
\displaystyle \Delta_{III} & \displaystyle\approx \frac{ft^{0-}}{ft^{0+}} \frac{ft^{1+}}{ft^{1-}} - (\delta_R'^{0+}-\delta_R'^{0-})+(\delta_R'^{1+}-\delta_R'^{1-}) \\[0.15in] 
& \displaystyle- (\delta_{NS}^{0+}-\delta_{NS}^{0-})+(\delta_{NS}^{1+}-\delta_{NS}^{1-}) + (\delta_C^{0+}-\delta_C^{0-})  \\[0.15in]
& \displaystyle + (\delta_C^{1+}-\delta_C^{1-}) - 1 . 
\end{array}
\end{equation}
Again, the residual $\Delta_{III}$ should vanish when appropriate theoretical and experimental inputs are provided. 
As an interesting feature of Eq.\,\eqref{test3}, any $Z$- and $Q$-dependent components embedded in the theoretical corrections would be almost completely canceled out, due to the double subtraction. Therefore, the term $(\delta_R'^{1+}-\delta_R'^{1-})-(\delta_R'^{0+}-\delta_R'^{0-})$ is expected to vanish. Regarding $\delta_{NS}$, besides the previously mentioned universal Born-graph contribution, it comprises an orbital isoscalar, spin isoscalar, orbital isovector, and spin isovector components\,\cite{TOWNER1973589,ToHa2002}. Both isoscalar components do not contribute to $\Delta_I$ and $\Delta_{III}$. However, the isovector components exhibit a sign reversal under the mirror symmetry, while maintaining a consistent magnitude. Consequently, the isovector contribution to $\Delta_{III}$ could be significant similar to $\Delta_{I}$, or negligible if substantial cancellation occurs between the lower and upper admixed states. A further detailed study of these radiative corrections would be highly necessary. 
\\
\indent
We have investigated the residual $\Delta_I$, $\Delta_{II}$ and $\Delta_{III}$ using the experimental $ft$ values from Ref.\,\cite{NNDCENSDF} and our shell-model results\,\cite{xayavong2023shellmodel}. 
Apart from the previously mentioned calculations of $\delta_{C1}$, the other leading-order term, $\delta_{C2}$, is evaluated with realistic Woods-Saxon radial wave functions. The potential depth and length parameter are readjusted to reproduce separation energies and charge radii whenever data are available. The uncertainty on $\delta_{C1}$ reflects the spread among different effective isospin-nonconserving interactions considered in the calculations. Meanwhile, the uncertainty on $\delta_{C2}$ arises from the variation due to different fitting methods and experimental errors in available charge-radius data. Further details on our uncertainty quantification can be found in Refs.\,\cite{xayavong2023shellmodel,XaNa2018}. The radiative corrections, $\delta_R'$ and $\delta_{NS}$ are not included due to the lack of numerical data. 
\\
\indent
Our results of experimental tests for the thirty-two pairs of mirror GT transitions, including upper admixed states, are illustrated in Fig.\,\ref{fig2}. The obtained $\Delta_I$ values are displayed in the top-left panel, and their individual uncertainty contributions in the top-right panel. The values of $\Delta_{II}$ and $\Delta_{III}$ are shown in the bottom-left and bottom-right panels, respectively. Among these three experimental tests, it is evident that $\Delta_I$ exhibits the greatest consistency. %, given that the large scatter points at $A=21$ are excluded. 
The corresponding $\chi^2/\nu$ value is 16.588. This large $\chi^2/\nu$ indicates a potential underestimate of the overall uncertainty and, possibly, the significance of the absent radiative corrections. Nevertheless, its weighted average is reasonably close to the ideal value, namely $\Delta_I=-0.131\pm0.110$. As indicated in the uncertainty budget provided in the top-right panel of Fig.\,\ref{fig2}, experimental errors predominantly dominate in most cases. In particular, we assume a zero uncertainty on $\delta_{C1}$ for nuclei resided in the $p-sd$ cross-shell region, due to the availability of only a single effective isospin-nonconserving interaction. Unexpectedly, the resulting values for the residual $\Delta_{II}$ deviate substantially from zero, with deviations largely exceeding the overall uncertainty, except for a few cases. Furthermore, the consistency of the $\Delta_{II}$ values cannot be maintained even if a conservative 100\,\% uncertainty is assumed for the isospin-symmetry breaking corrections. 
This strongly suggests the significance of the missing radiative corrections or, more likely, the unreliability of the shell-model values of $\eta^2$. The arithmetic mean of $\Delta_{II}$ is $0.890\pm4.028$. However, it is interesting to remark that the $\Delta_{II}$ values for a given pair of mirror GT transitions are nearly coincident. This observation seems to indicate that the missing radiative corrections are nearly invariant under the mirror symmetry as expected, despite the significance of their individual values. 
\\
\indent
The last experimental test of our theoretical calculations is based on the residual $\Delta_{III}$. From the results displayed in the bottom-right panel of Fig.\,\ref{fig2}, $\Delta_{III}$ values are much closer to zero compared with those of $\Delta_{II}$, due to the mirror symmetry property of the missing radiative corrections, as remarked above. Nevertheless, it is likely that the overall uncertainties of $\Delta_{III}$ are underestimated for most cases, except for $A=17,22,25$ and $31$. Excluding these exceptional cases, this test yields an extremely large $\chi^2/\nu$ value. The arithmetic mean of $\Delta_{III}$ is 0.154$\pm$0.940. 
\\
\indent
{\it Conclusion:} To summarize, the boundaries\,\eqref{con}, derived from the simplified two-level isospin-mixing model, define permissible regions in the ($\eta,\delta_{C1}$) plane. This serves as a filter for theoretical calculations, particularly when $\eta$ approaches zero. Our current shell-model results successfully satisfy this criterion for all transitions, including $A=22$ where an exceptionally strong isospin mixing has been observed. Regarding the experimental tests, the remaining deviations of $\Delta_I$ and $\Delta_{III}$ from zero seems to be principally due to inaccuracy in the our theoretical calculations of the isospin-symmetry breaking correction and errors in the experimental data. The contributions of the radiative corrections to these residuals are expected to be negligible owing to the mirror symmetry, especially for $\Delta_{III}$. Conversely, $\Delta_{II}$ is substantially deteriorated as the result of the unreliability of the shell-model prediction of $\eta^2$, and possibly the absence of the radiative corrections. As a byproduct, the incorporation of the upper admixed states enable the experimental extraction of $\eta^2$. This serves as a test exclusively sensitive to the isospin-symmetry part of the nuclear Hamiltonian. Remarkably, this test was found to be meaningful in several cases, even if a conservative 50\,\% uncertainty is assumed to cover the corrective contributions. 
\\
\indent
\begin{acknowledgments} 
L.\,Xayavong and Y.\,Lim are supported by the National Research Foundation of Korea(NRF) grant funded by the Korea government(MSIT)(No. 2021R1A2C2094378). Y.\,Lim is also supported by the Yonsei University Research Fund of 2023-22-0126. 
%\added{N. A. Smirnova acknowledges the IN2P3/CNRS, France, in the framework of the ``Isospin-symmetry breaking'' and ``Exotic nuclei, fundamental interactions and astrophysics'' Master projects}. 
\end{acknowledgments}
%


%\bibliography{2lev_model}

%\clearpage %% for a page break
\appendix
\section{Supplementary material}
%Supplementary material
%\appendix{Supplementary material}
\begin{table*}%[h!]
\caption{Numerical results of the shell-model calculations for the ratio $\eta$ and the total isospin-symmetry breaking correction. The + (-) label indicates the $\beta^+$ ($\beta^-$) decays. The 0 (1) label indicates the lower (upper) admixed states. The unit of the correction is \%. }\label{tab01}
\begin{ruledtabular}
\begin{tabular}{c|c|c|c|c|c|c|c} 
Mirror pairs 	&	 $J_i^\pi T_i$ 	&	 $J_f^\pi T_f$ 	&	$\eta$	&	$\delta_C^{0-}$	&	$\delta_C^{0+}$	&	$\delta_C^{1-}$	&	$\delta_C^{1+}$	\\
\hline				
$^8$Li$(\beta^-)^8$Be; $^8$B($\beta^+$)$^8$Be 	&	 $2^+ 1$ 	&	 $2_1^+ 0$ 	&	7.514$\pm$1.812	&	-7.924$\pm$2.647	&	7.440$\pm$2.074	&	4.632$\pm$1.788	&	6.766$\pm$1.604	\\
$^{12}$B($\beta^-$)$^{12}$C; $^{12}$N($\beta^+$)$^{12}$C 	&	 $1^+ 1$ 	&	 $0_1^+ 0$ 	&	1.929$\pm$0.074	&	-4.976$\pm$0.636	&	11.530$\pm$0.840	&	0.569$\pm$0.638	&	3.117$\pm$0.904	\\
$^{13}$B($\beta^-$)$^{13}$C; $^{13}$O($\beta^+$)$^{13}$N 	&	$\frac{3}{2}^- \frac{3}{2}$ 	&	$\frac{1}{2}_1^- \frac{1}{2}$ 	&	0.332$\pm$0.481	&	-2.981$\pm$0.955	&	4.484$\pm$0.221	&	6.428$\pm$0.421	&	9.880$\pm$0.356	\\
	&	$\frac{3}{2}^- \frac{3}{2}$ 	&	$\frac{3}{2}_1^- \frac{1}{2}$ 	&	2.102$\pm$0.000	&	-8.281$\pm$0.278	&	5.327$\pm$1.330	&	-6.123$\pm$0.598	&	2.507$\pm$0.740	\\
$^{17}$N($\beta^-$)$^{17}$O; $^{17}$Ne($\beta^+$)$^{17}$F 	&	$\frac{1}{2}^- \frac{3}{2}$ 	&	$\frac{3}{2}_1^- \frac{1}{2}$ 	&	3.530$\pm$0.000	&	13.318$\pm$0.632	&	44.670$\pm$2.596	&	-1.026$\pm$0.286	&	-13.324$\pm$0.442	\\
	&	$\frac{1}{2}^- \frac{3}{2}$ 	&	$\frac{1}{2}_1^- \frac{1}{2}$ 	&	3.525$\pm$0.000	&	11.261$\pm$0.033	&	3.579$\pm$0.120	&	3.137$\pm$0.112	&	5.838$\pm$0.106	\\
$^{20}$F($\beta^-$)$^{20}$Ne; $^{20}$Na($\beta^+$)$^{20}$Ne 	&	 $2^+ 1$ 	&	 $2_1^+ 0$ 	&	1.167$\pm$0.000	&	6.169$\pm$0.211	&	14.201$\pm$0.636	&	20.374$\pm$0.130	&	20.164$\pm$0.246	\\
$^{20}$O($\beta^-$)$^{20}$F; $^{20}$Mg($\beta^+$)$^{20}$Na 	&	 $0^+ 2$ 	&	 $1_1^+ 1$ 	&	0.742$\pm$0.000	&	8.940$\pm$0.136	&	12.402$\pm$0.077	&	-1.071$\pm$0.021	&	-11.096$\pm$0.526	\\
$^{21}$F($\beta^-$)$^{21}$Ne; $^{21}$Mg($\beta^+$)$^{21}$Na 	&	$\frac{5}{2}^+ \frac{3}{2}$ 	&	$\frac{3}{2}_1^+ \frac{1}{2}$ 	&	2.341$\pm$0.000	&	5.879$\pm$0.122	&	10.650$\pm$0.610	&	18.342$\pm$0.206	&	22.311$\pm$0.412	\\
	&	$\frac{5}{2}^+ \frac{3}{2}$ 	&	$\frac{5}{2}_1^+ \frac{1}{2}$ 	&	0.131$\pm$0.000	&	6.145$\pm$0.025	&	11.068$\pm$0.045	&	-256.639$\pm$0.349	&	-417.041$\pm$0.652	\\
$^{22}$O($\beta^-$)$^{22}$F; $^{22}$Si($\beta^+$)$^{22}$Al 	&	 $0^+ 3$ 	&	 $1_1^+ 2$ 	&	7.665$\pm$0.000	&	96.917$\pm$0.283	&	88.616$\pm$	0.135	&	79.224$\pm$0.158	&	87.694$\pm$0.050	\\
$^{24}$Ne($\beta^-$)$^{24}$Na; $^{24}$Si($\beta^+$)$^{24}$Al 	&	 $0^+ 2$ 	&	 $1_1^+ 1$ 	&	1.725$\pm$0.000	&	0.958$\pm$0.428	&	21.225$\pm$0.497	&	0.860$\pm$0.020	&	14.754$\pm$0.033	\\
$^{25}$Na($\beta^-$)$^{25}$Mg; $^{25}$Si($\beta^+$)$^{25}$Al 	&	$\frac{5}{2}^+ \frac{3}{2}$ 	&	$\frac{5}{2}_1^+ \frac{1}{2}$ 	&	0.334$\pm$0.460	&	1.958$\pm$0.717	&	3.072$\pm$2.535	&	9.610$\pm$0.428	&	6.632$\pm$0.181	\\
	&	$\frac{5}{2}^+ \frac{3}{2}$ 	&	$\frac{3}{2}_1^+ \frac{1}{2}$ 	&	0.956$\pm$0.000	&	-0.523$\pm$0.013	&	9.335$\pm$0.082	&	0.396$\pm$0.011	&	0.239$\pm$0.207	\\
$^{26}$Na($\beta^-$)$^{26}$Mg; $^{26}$P($\beta^+$)$^{26}$Si 	&	 $3^+ 2$ 	&	 $2_1^+ 1$ 	&	0.079$\pm$0.023	&	4.278$\pm$0.159	&	13.040$\pm$1.958	&	-50.555$\pm$1.185	&	61.994$\pm$2.848	\\
$^{31}$Al($\beta^-$)$^{31}$Si; $^{31}$Ar($\beta^+$)$^{31}$Cl 	&	$\frac{5}{2}^+ \frac{5}{2}$ 	&	$\frac{3}{2}_1^+ \frac{3}{2}$ 	&	1.336$\pm$1.543	&	-1.518$\pm$0.251	&	9.089$\pm$2.899	&	-9.307$\pm$1.248	&	-7.650$\pm$0.624	\\
%\hline	
\end{tabular}
\end{ruledtabular}
\end{table*}

\begin{table*}[h]
\caption{Experimental data for $\log ft$\,\cite{NNDCENSDF}. The + (-) label indicates the $\beta^+$ ($\beta^-$) decays. The 0 (1) label indicates the lower (upper) admixed states.
}\label{tab2}
\begin{ruledtabular}
\begin{tabular}{c|c|c|c|c|c|c} 
Mirror Pair &	 $J_i^\pi T_i$ 	&	 $J_f^\pi T_f$ 	&	$\log ft^{0-}$	&	$\log ft^{1-}$	&	$\log ft^{0+}$	&	$\log ft^{1+}$	\\
\hline	
$^8$Li$(\beta^-)^8$Be; $^8$B($\beta^+$)$^8$Be	&	 $2^+ 1$ 	&	 $2_1^+ 0$ 	&	5.589$\pm$0.008	&		&	5.622$\pm$0.008	&	3.01$\pm$0.05	\\
$^{12}$B($\beta^-$)$^{12}$C; $^{12}$N($\beta^+$)$^{12}$C	&	 $1^+ 1$ 	&	 $0_1^+ 0$ 	&	4.0617$\pm$0.0005	&	4.572$\pm$0.017	&	4.1106$\pm$0.0007	&	4.622$\pm$0.01	\\
$^{13}$B($\beta^-$)$^{13}$C; $^{13}$O($\beta^+$)$^{13}$N	&	$\frac{3}{2}^- \frac{3}{2}$ 	&	$\frac{1}{2}_1^- \frac{1}{2}$ 	&	4.034$\pm$0.006	&	4.59$\pm$0.09	&	4.081$\pm$0.011	&	4.66$\pm$0.1	\\
	&	$\frac{3}{2}^- \frac{3}{2}$ 	&	$\frac{3}{2}_1^- \frac{1}{2}$ 	&	4.45$\pm$0.05	&	4.95$\pm$0.2	&	4.55$\pm$0.01	&	5.09$\pm$0.15	\\
$^{17}$N($\beta^-$)$^{17}$O; $^{17}$Ne($\beta^+$)$^{17}$F	&	$\frac{1}{2}^- \frac{3}{2}$ 	&	$\frac{3}{2}_1^- \frac{1}{2}$ 	&	4.416$\pm$0.015	&	3.851$\pm$0.013	&	4.65$\pm$0.03	&	3.895$\pm$0.024	\\
	&	$\frac{1}{2}^- \frac{3}{2}$ 	&	$\frac{1}{2}_1^- \frac{1}{2}$ 	&	7.08$\pm$0.09	&	4.37$\pm$0.04	&	7.13$\pm$0.19	&	4.55$\pm$0.02	\\
$^{20}$F($\beta^-$)$^{20}$Ne; $^{20}$Na($\beta^+$)$^{20}$Ne	&	 $2^+ 1$ 	&	 $2_1^+ 0$ 	&	4.9788$\pm$0.0003	&			&	4.987$\pm$0.003	&	4.196$\pm$0.007	\\
$^{20}$O($\beta^-$)$^{20}$F 	&	 $0^+ 2$ 	&	 $1_1^+ 1$ 	&	3.734$\pm$0.0006	&	3.64$\pm$0.06	&	3.81$\pm$0.03	&	4.06$\pm$0.07	\\
$^{21}$F($\beta^-$)$^{21}$Ne; $^{21}$Mg($\beta^+$)$^{21}$Na	&	$\frac{5}{2}^+ \frac{3}{2}$ 	&	$\frac{3}{2}_1^+ \frac{1}{2}$ 	&	5.67$\pm$0.14	&	4.52$\pm$0.03	&	5.26$\pm$0.13	&	4.48$\pm$0.03	\\
	&	$\frac{5}{2}^+ \frac{3}{2}$ 	&	$\frac{5}{2}_1^+ \frac{1}{2}$ 	&	4.662$\pm$0.002	&	7.1$\pm$0.6	&	7.72$\pm$0.27	&	5.9$\pm$0.17	\\
$^{22}$O($\beta^-$)$^{22}$F; $^{22}$Si($\beta^+$)$^{22}$Al 	&	 $0^+ 3$ 	&	 $1_1^+ 2$ 	&	4.6$\pm$0.1	&	3.8$\pm$0.1	&	5.09$\pm$0.09	&	3.83$\pm$0.05	\\
$^{24}$Ne($\beta^-$)$^{24}$Na; $^{24}$Si($\beta^+$)$^{24}$Al	&	 $0^+ 2$ 	&	 $1_1^+ 1$ 	&	4.364$\pm$0.003	&	4.4$\pm$0.012	&	4.36$\pm$0.05	&	4.45$\pm$0.03	\\
$^{25}$Na($\beta^-$)$^{25}$Mg; $^{25}$Si($\beta^+$)$^{25}$Al	&	$\frac{5}{2}^+ \frac{3}{2}$ 	&	$\frac{5}{2}_1^+ \frac{1}{2}$ 	&	5.25$\pm$0.02	&	6.04$\pm$0.1	&	5.24$\pm$0.14	&	6.21$\pm$0.13	\\
	&	$\frac{5}{2}^+ \frac{3}{2}$ 	&	$\frac{3}{2}_1^+ \frac{1}{2}$ 	&	5.05$\pm$0.03	&	5.19$\pm$0.08	&	5.05$\pm$0.08	&	5.43$\pm$0.03	\\
$^{26}$Na($\beta^-$)$^{26}$Mg; $^{26}$P($\beta^+$)$^{26}$Si	&	 $3^+ 2$ 	&	 $2_1^+ 1$ 	&	4.71$\pm$0.01	&	7.6$\pm$0.4	&	4.9$\pm$0.1	&	5.9$\pm$0.3	\\
$^{31}$Al($\beta^-$)$^{31}$Si; $^{31}$Ar($\beta^+$)$^{31}$Cl	&	$\frac{5}{2}^+ \frac{5}{2}$ 	&	$\frac{3}{2}_1^+ \frac{3}{2}$ 	&	4.77$\pm$0.06	&	4.47$\pm$0.13	&	4.93$\pm$0.02	&	4.55$\pm$0.06	\\
\end{tabular}
\end{ruledtabular}
\end{table*}

\begin{table*}[h]
\caption{Numerical results for the residuals defined in Eq.\,\eqref{test1}, \eqref{test2}, and \eqref{test3}. The + (-) label indicates the $\beta^+$ ($\beta^-$) decays. The 0 (1) label indicates the lower (upper) admixed states. 
}\label{tab3}
\begin{ruledtabular}
\begin{tabular}{c|c|c|c|c|c|c|c} 
Mirror pairs 	&	 $J_i^\pi T_i$ 	&	 $J_f^\pi T_f$ 	&	$\Delta_I^0$	&	$\Delta_I^1$	&	$\Delta_{II}^+$		&	$\Delta_{II}^-$	&	$\Delta_{III}$	\\
\hline				
$^8$Li$(\beta^-)^8$Be; $^8$B($\beta^+$)$^8$Be 	&	 $2^+ 1$ 	&	 $2_1^+ 0$ 	&	-0.075$\pm$0.036	&		&	-350.048$\pm$29.453	&		&		\\
$^{12}$B($\beta^-$)$^{12}$C; $^{12}$N($\beta^+$)$^{12}$C 	&	 $1^+ 1$ 	&	 $0_1^+ 0$ 	&	-0.046$\pm$0.011	&	0.097$\pm$0.025	&	3.440$\pm$0.285	&	3.396$\pm$0.285	&	0.193$\pm$0.018	\\
$^{13}$B($\beta^-$)$^{13}$C; $^{13}$O($\beta^+$)$^{13}$N 	&	$\frac{3}{2}^- \frac{3}{2}$ 	&	$\frac{1}{2}_1^- \frac{1}{2}$ 	&	0.040$\pm$0.017	&	0.140$\pm$0.158	&	-0.168$\pm$0.320	&	-0.194$\pm$0.320	&	0.164$\pm$0.108	\\
	&	$\frac{3}{2}^- \frac{3}{2}$ 	&	$\frac{3}{2}_1^- \frac{1}{2}$ 	&	0.123$\pm$0.066	&	0.294$\pm$0.345	&	4.139$\pm$0.005	&	4.096$\pm$0.016	&	0.319$\pm$0.175	\\
$^{17}$N($\beta^-$)$^{17}$O; $^{17}$Ne($\beta^+$)$^{17}$F 	&	$\frac{1}{2}^- \frac{3}{2}$ 	&	$\frac{3}{2}_1^- \frac{1}{2}$ 	&	0.400$\pm$0.063	&	0.230$\pm$0.031	&	10.070$\pm$0.166	&	9.314$\pm$0.054	&	-0.164$\pm$25.412	\\
	&	$\frac{1}{2}^- \frac{3}{2}$ 	&	$\frac{1}{2}_1^- \frac{1}{2}$ 	&	0.199$\pm$0.236	&	0.487$\pm$0.068	&	-376.352$\pm$73.871	&	-458.765$\pm$42.412	&	0.299$\pm$0.285	\\
$^{20}$F($\beta^-$)$^{20}$Ne; $^{20}$Na($\beta^+$)$^{20}$Ne 	&	 $2^+ 1$ 	&	 $2_1^+ 0$ 	&	-0.061$\pm$0.007	&		&	-5.188$\pm$0.046	&		&		\\
$^{20}$O($\beta^-$)$^{20}$F; $^{20}$Mg($\beta^+$)$^{20}$Na 	&	 $0^+ 2$ 	&	 $1_1^+ 1$ 	&	0.157$\pm$0.036	&	1.731$\pm$0.243	&	0.121$\pm$0.013	&	-0.566$\pm$0.002	&	1.142$\pm$0.168	\\
$^{21}$F($\beta^-$)$^{21}$Ne; $^{21}$Mg($\beta^+$)$^{21}$Na 	&	$\frac{5}{2}^+ \frac{3}{2}$ 	&	$\frac{3}{2}_1^+ \frac{1}{2}$ 	&	-0.659$\pm$0.075	&	-0.128$\pm$0.039	&	-1.249$\pm$0.876	&	-10.407$\pm$2.224	&	1.432$\pm$0.453	\\
	&	$\frac{5}{2}^+ \frac{3}{2}$ 	&	$\frac{5}{2}_1^+ \frac{1}{2}$ 	&	1,141.829$\pm$308.586	&	0.667$\pm$0.040	&	216.797$\pm$58.532	&	0.023$\pm$0.000	&	-2.555$\pm$0.007	\\
$^{22}$O($\beta^-$)$^{22}$F; $^{22}$Si($\beta^+$)$^{22}$Al 	&	 $0^+ 3$ 	&	 $1_1^+ 2$ 	&	2.173$\pm$0.416	&	-0.013$\pm$0.120	&	40.728$\pm$1.623	&	53.564$\pm$0.520	&	-0.652$\pm$48,769.831	\\
$^{24}$Ne($\beta^-$)$^{24}$Na; $^{24}$Si($\beta^+$)$^{24}$Al 	&	 $0^+ 2$ 	&	 $1_1^+ 1$ 	&	-0.212$\pm$0.050	&	-0.017$\pm$0.036	&	2.216$\pm$0.038	&	2.057$\pm$0.005	&	0.474$\pm$0.067	\\
$^{25}$Na($\beta^-$)$^{25}$Mg; $^{25}$Si($\beta^+$)$^{25}$Al 	&	$\frac{5}{2}^+ \frac{3}{2}$ 	&	$\frac{5}{2}_1^+ \frac{1}{2}$ 	&	-0.034$\pm$0.141	&	0.509$\pm$0.243	&	0.000$\pm$0.308	&	-0.063$\pm$0.307	&	0.495$\pm$0.292	\\
	&	$\frac{5}{2}^+ \frac{3}{2}$ 	&	$\frac{3}{2}_1^+ \frac{1}{2}$ 	&	-0.099$\pm$0.085	&	0.739$\pm$0.148	&	0.534$\pm$0.030	&	0.182$\pm$0.022	&	0.835$\pm$7.415	\\
$^{26}$Na($\beta^-$)$^{26}$Mg; $^{26}$P($\beta^+$)$^{26}$Si 	&	 $3^+ 2$ 	&	 $2_1^+ 1$ 	&	0.461$\pm$0.157	&	-2.106$\pm$0.032	&	-0.143$\pm$0.016	&	0.006$\pm$0.004	&	0.226$\pm$0.037	\\
$^{31}$Al($\beta^-$)$^{31}$Si; $^{31}$Ar($\beta^+$)$^{31}$Cl 	&	$\frac{5}{2}^+ \frac{5}{2}$ 	&	$\frac{3}{2}_1^+ \frac{3}{2}$ 	&	0.339$\pm$0.096	&	0.186$\pm$0.173	&	-0.212$\pm$4.124	&	-0.055$\pm$4.125	&	-0.046$\pm$10,047.546	\\
\end{tabular}
\end{ruledtabular}
\end{table*}

\end{document}